\documentstyle{article}
\font \msb=msbm10 scaled \magstep1
\newcommand{\bR}{\mbox{\msb R} }
\newcommand{\bC}{\mbox{\msb C} }
\newcommand{\bP}{\mbox{\msb P} }

\font \eul=eufm10 scaled \magstep2

\newcommand{\sm}{\mbox{\eul s}}

\def\a{{\alpha}}
\def\b{{\beta}}
\def\g{{\gamma}}
\def\d{{\delta}}

\def\l{{\lambda}}
\def\m{{\mu}}
\def\s{{\sigma}}

\def\w{{ \Omega }}

\def\p{\partial}

\newcommand{\bs}{\mbox{\bf S}}
\newcommand{\bc}{\mbox{\bf C}}
\newcommand{\gm}{\mbox{\cal g}}
\def\A{{\cal A}}
\def\B{{\cal B}}
\def\C{{\cal C}}
\def\DD{{\cal D}}
\def\H{{\cal H}}
\def\HH{{\sf H}}
\def\L{{\cal L}}
\def\M{{\cal M}}
\def\U{{\cal U}}
\def\sx{{\sl x}}
\def\sp{{\sl p }}
\def\te#1{{\widetilde{#1}}}

\def\on#1#2{\mathop{\vbox{\ialign{##\crcr\noalign{\kern2pt}
$\scriptstyle{#2}$\crcr\noalign{\kern2pt\nointerlineskip}
\kern-2pt$\hfil\displaystyle{#1}\hfil$\crcr}}}\limits}

\def\nn{ \nonumber }
\def\bq{ \begin{equation} }
\def\eq{ \end{equation} }
\def\ben{ \begin{eqnarray} }
\def\en{ \end{eqnarray} }
\def\ll{ \label }
\def\dfrac#1#2{{\displaystyle{#1\over#2}}}
\def\frac#1#2{{{#1\over#2}}}

\newtheorem{th}{Theorem}
\newtheorem{prop}{Proposition}

\begin{document}

\title{The St\"{a}ckel systems and algebraic curves.}
\author{
 A.V. Tsiganov\\
{\small\it
 Department of Mathematical and Computational Physics,
 Institute of Physics,}\\
{\small\it
St.Petersburg University,
198 904,  St.Petersburg,  Russia}
}
\date{}
\maketitle

\begin{abstract}
We show how the Abel-Jacobi map provides all the principal
properties of an ample family of integrable mechanical systems
associated to hyperelliptic curves.  We prove that
derivative of the Abel-Jacobi map is just the St\"{a}ckel
matrix, which determines $n$-orthogonal curvilinear  coordinate
systems in a flat space. The Lax pairs, $r$-matrix algebras and
explicit form of the flat coordinates are constructed.  An
application  of the Weierstrass reduction theory allows to
construct several flat coordinate systems on a common
hyperelliptic curve and to connect among themselves different
integrable systems on a single phase space.
\end{abstract}

\vskip 0.5cm
\begin{flushleft}
\sf arch-ive/9712003
\end{flushleft}
\vskip 1cm

\section{Introduction}
\setcounter{equation}{0}
In the classical mechanics  the arrow from the initial physical
variables to the action-angle variables is provided by the
separation of variables and then by the Arnold construction of
the action-angle representation \cite{arn89}. The motion in the
opposite direction ought to allow us to construct various
mechanical integrable systems.  However, in the action-angle
representation all the mechanical systems with fixed number of
degrees of freedom are indistinguishable.  To describe some
particular integrable system one should present an explicit
construction of the initial physical variables as functions on
the action-angle variables.  This mapping contains all the
information about a given integrable system. By using variety
of these mappings the different integrable models may be
connected together via the common action-angle variables. For
instance, mechanical systems may be related to nonlinear
equations and to gauge field theory.

As an example, investigation of the finite-gap solutions of the
nonlinear problems leads to the introduction \cite{nov82,du81,krp96}
of analytic symplectic form $\w_g$ on the Jacobian fibrations
and to the definition of the action-angle variables on the
complex space of Liouville variables.  In \cite{neh72} it is
shown that possible obstructions to the existence of global
systems of action-angle variables on symplectic vector bundles
are a nontrivial first Chern class and the presence of
monodromy at singularities. Introduction of the action-angle
representation enables ones to consider mechanical integrable
systems as systems associated with  these variables on a torus
bundle with base $\M$, moduli space of complex polynomials
\bq
F(\l)=\prod_{j=1}^{2g+1}(\l-\l_j)\,,\ll{bac}
\eq
and with a fiber $J(\C)$, the
$g$-dimensional complex Jacobian of auxiliary curve $\C$
defined by the Abel-Jacobi map $\U$ \cite{arn89,fr87}.  The
fact that action-angle variables could be used for quantization
of classical systems leads to introduction of semiclassical
geometric phases. This approach results, for instance, in a
quantum conditions on the moduli of $n$-dimensional Jacobi
varieties \cite{ber85}.

By using this Abel-Jacobi map $\U$ and the Jacobi problem of
inversion, the so-called root variables $\{p_j,q_j\}_{j=1}^n$
\cite{al81,almar92} on an associated Riemann surface $\C$
may be  constructed instead of the action-angle variables.  In
these root variables  on the level of integrals of motion the
action is represented as a sum of items depending on one
coordinate only, i.e. these variables are separated variables.
The corresponding Riemann surface $\C$ depends on parameters
(moduli), parameterizing the moduli space $\M$ of $\C$
\cite{krp96,man96}. In terms of mechanical integrable systems
the curve $\C$ is interpreted as a time-independent spectral
curve, integrals of motion are some specific coordinates on the
moduli space $\M$ and Jacobian $J(\C)$ is a common level of the
involutive integrals of the system \cite{arn89}.

In what follows, we have to describe appropriate mechanical
systems together with their phase space in initial physical
coordinates $\{\sp_j,\sx_j\}_{j=1}^n$.  In particular, separated
coordinate systems ought to be orthogonal curvilinear
coordinate systems on the flat Riemannian manifold
\cite{eis34,vil68}.  In this case, these separated coordinate
systems are associated to some solutions to the Lam\'{e}
equation \cite{zah96,kr96,razs96}.  Recently, the solutions to
this equation have been obtained in an explicit form  with
the help of the "dressing procedure" \cite{zah96}, the
Baker-Akhiezer function \cite{kr96,kr94} and the Lie algebraic
construction \cite{razs96} within framework of the inverse
problem method.

The main objective this paper is to illustrate how fixed
mapping from the action-angle variables \cite{arn89,fr87,kr96}
to separated variables completely defines all the principal
properties of mechanical systems.
We shall consider the
uniform St\"{a}ckel models associated to the Abel-Jacobi map
$\U$ on the hyperelliptic curve $\C$ and the well-known
elliptic, parabolic and spherical curvilinear coordinate
systems on $\bR^n$. Also we discuss relations of these
mechanical systems with other integrable models associated to
the same algebraic curve.


\section{The St\"{a}ckel systems}
\setcounter{equation}{0}
One of the oldest problem of the hamiltonian mechanics is to
find the quadratures for the integrable hamiltonian systems.
The simplest models integrable in quadratures are the Liouville
systems and the St\"{a}ckel systems \cite{st95} ( the Liouville
systems are a particular case of the St\"{a}ckel systems).

Before proceeding father it is useful to recall the classical
work of St\"{a}ckel \cite{st95}.  The system associated with
the name of St\"{a}ckel \cite{st95} is a holonomic system on
the phase space $\bR^{2n}$, their hamiltonian is
\bq
H=\sum_{j=1}^n \gm_{j}(q_1,\ldots,q_n)
\left(p_j^2+U_j\right)\,.\ll{sth}
\eq
Here $\{p_j,q_j\}_{j=1}^n$ are canonical variables in $\bR^{2n}$
with the standard symplectic structure and with the following
Poisson brackets
\bq \w_n=\sum_{j=1}^n dp_j\wedge dq_j\,,\qquad
\{p_j,q_k\}=\d_{jk}\,.\ll{stw}
\eq
There is an even stronger version of the St\"ackel theorem.
\begin{th}
For a hamiltonian system with hamiltonian H of the form (\ref{sth})
the following assertions are equivalent:
\par 1) The associated Hamilton-Jacobi equation is separable.
\par 2)
There exists a nondegenerate $n\times n$
St\"{a}ckel matrix $\bs$, whose elements $\sm_{kj}$ depend
only on $q_j$
\ben
&& \det \bs\neq 0\,,\qquad  \dfrac{\p\sm_{kj}}{\p q_m}=0\,,
\quad{\rm for}\quad j\neq m\nn\\
{\rm~and~such~that}\qquad&&\ll{stm}\\
&&
\sum_{j=1}^n \sm_{kj}(q_j) {\rm \gm}_{j}(q_1,\ldots,q_n)=\d_{k1}\,.\nn
\en
\par 3) There exist $n$ functionally independent integrals of motion
which are quadratic in momenta.
\end{th}
Let $\bc=[c_{ik}]$ denotes inverse matrix to $\bs$ such that
$c_{j1}=\gm_{j}$. Then the St\"{a}ckel theorem \cite{st95,eis34}
asserts that there are $n$ first integrals of motion, namely
\bq
I_k=\sum_{j=1}^n c_{jk}\left(p_j^2+U_j\right)\,,\qquad I_1=H\,.
\ll{fint}
\eq
The common level surface of these integrals
\[M_\a=\left\{z\in \bR^{2n}: I_k(z)=\a_k\,,~k=1,\ldots,n\right\}\]
is diffeomorphic to the $n$-dimensional real torus
and  one immediately gets
\bq
p_j^2=\left(\dfrac{\p S}{\p q_j}\right)^2=
\sum_{k=1}^n \a_k\sm_{kj}(q_j)-U_j(q_j)\,,\ll{stc}
\eq
where $S(q_1\,\ldots,q_n)$ is an action function
\cite{arn89}.  If this real torus is a part of complex
algebraic torus, then the corresponding mechanical system is
called an algebraic completely integrable system \cite{avm89}.

The St\"{a}ckel theorem allows to reduce the solution of
the equations of motion to a problem in algebraic geometry.
We can regard each expression (\ref{stc}) as being defined on
the Riemann surface
\bq
\C_j:\quad
y_j^2=F_j(\l)\,,\qquad F_j(\l)=\sum_{k=1}^n \a_k\sm_{kj}(\l)-U_j(\l)\,,
\ll{sthc}
\eq
which depends on the values $\a_k$ of integrals of motion.
All the pairs of variables $(p_j,q_j)$ lie on these Riemann
surfaces (\ref{sthc}).  Considered together, they determine an
$n$-dimensional Lagrangian submanifold in $\bR^{2n}$
\bq
\C^{(n)}:\qquad \C_1(p_1,q_1)\times\C_2(p_2,q_2)
\times\cdots\times\C_n(p_n,q_n)\,.\ll{stl}\eq
The associated Hamilton-Jacobi equation
\bq
\dfrac{\p S}{\p t}+
H(t,\dfrac{\p S}{\p q_1},\ldots,
\dfrac{\p S}{\p q_n},q_1,\ldots,q_n)=0\,,\qquad\Rightarrow\qquad
\gm^{jj}\,\dfrac{\p S}{\p q_j}\,
\dfrac{\p S}{\p q_j}=E\,,\ll{hjeq}
\eq
on the local manifold ${\cal V}_n$ with diagonal metric
$\gm^{jj}=\gm_j(q_1,\ldots,q_n)$ analytic in the local
coordinates $\{q_j\}$ has the following additive solution
\bq
S(q_1\,\ldots,q_n)=\sum_{j=1}^n s_j(q_j)\,,\qquad
s_j(q_j)=\int{\sqrt{F_j(q_j)~~}~d q_j}\,,\ll{stac}
\eq
with the functions $F_j(\l)$ defined in (\ref{sthc}).
Coordinates $q_j(t,\a_1,\ldots,\a_n)$ are determined from the equations
\ben
\sum_{j=1}^n\int_{\g_{0}(p_0,q_0)}^{\g_j(p_j,q_j)}\dfrac{\sm_{1j}(\l) d\l}
{\sqrt{\sum_{k=1}^n \a_k\sm_{1j}(\l)-U_j(\l)}}&=&\b_1=t\,,\nn\\
\ll{stinv}\\
\sum_{j=1}^n\int_{\g_{0}(p_0,q_0)}^{\g_j(p_j,q_j)}\dfrac{\sm_{kj}(\l) d\l}
{\sqrt{\sum_{k=1}^n \a_k\sm_{kj}(\l)-U_j(\l)}}&=&\b_k\,,
\qquad k=2,\ldots,n\,,\nn
\en
where points $\g_j(p_j,q_j)$ and $\g_{0}(p_0,q_0)$ be on the
curve $\C_j$ (\ref{sthc}).  Notice, that bounded motion in this
case will not be periodic in general but only  conditionally
periodic \cite{st95,eis34}.  If $\l_{j}^0$ and $\l_j$ are the
turning points determined by the conditions that functions
$F_j(\l)$ (\ref{sthc},\ref{stac}) vanish, the periods of the
motion $w_{jk}$  are equal to
\bq
w_{kj}=\int_{\l^0_j}^{\l_j}
\dfrac{ \sm_{kj}(\l)d\l }{ \sqrt{F_j(\l)} }\,.
\ll{stp}
\eq
Thus, St\"{a}ckel \cite{st95} showed that the orthogonal
coordinates $\{q_j\}_{j=1}^n$ permit separation in the
Hamilton-Jacobi equation (\ref{hjeq}) if the metric
\bq
ds^2=\sum_{j=1}^n \gm_{jj}(q_1,\ldots,q_n)\,(dq^j)^2\,,\qquad
\gm_{jj}(q_1,\ldots,q_n)\equiv \gm_{j}(q_1,\ldots,q_n)
\ll{metr}
\eq
is in the St\"{a}ckel form
\bq
\gm_{jj}(q_1,\ldots,q_n)=\HH_j^2(q_1,\ldots,q_n)
=\dfrac{\det\bs}{\bs^{j1}}\,,\ll{stlam}
\eq
where $\bs^{j1}$ means the cofactor of $\sm_{j1}$ in matrix
$\bs$ (\ref{stm}).  Here  $\gm_{jj}$ is a diagonal metric and
$\HH_j$ are called the Lam\'{e} coefficients. The
modern approach to construction of the Lam\'e coefficients
see in \cite{zah96,kr96,razs96}.

Henceforth, we shall restrict our attention to the uniform
St\"{a}ckel systems, where all the potentials $U_j(q_j)=
U(q_j)$ and curves $\C_j$ (\ref{sthc}) are equal.
Variables $\{s_k,w_k\}$ (\ref{stac},\ref{stp}) on a single curve
$\C$ are the action-angle variables for the uniform St\"{a}ckel
systems.  To construct the metric $\gm_j(q_1,\ldots,q_n)$ and
the potentials $U(q_j)$ in an explicit form we shall identify
periods $w_{k}$ (\ref{stp}) with  periods of the Abel
differentials on a common hyperelliptic curve $\C$ (\ref{stc}) along the
elements of a homology basis \cite{arn89,fr87}. In this case
definition of the separated variables $\{q_j\}$ (\ref{stinv}) leads to
the Jacobi inversion problem.  In the next Section we prove
that the St\"{a}ckel matrix $\bs$ (\ref{stm}) is completely
defined by the derivative of the Abel-Jacobi map $\U$ on $\C$
at generic point (so-called Brill-Noether matrix).


\section{Uniform St\"{a}ckel systems and algebraic curves}
\setcounter{equation}{0}
To begin with let us briefly recall some necessary facts about
the action-angle variables on the Jacobian $J(\C)$
\cite{arn89,fr87,krp96}. The main ingredient of this construction
is a universal configuration space, which is the moduli space
\cite{man96} of all algebraic curves with fixed jets of local
coordinates at a fixed number of punctures. This concept is
closely related to the notion of the Baker-Akhiezer function on
admissible curves \cite{kr96} and to the theory of algebraic
completely integrable systems \cite{avm89}.

Let us consider a genus $g$ Riemann surface $\C$ with
$N$ ordered punctures $P_j$ and with two special Abelian
integrals $y$ and $\l$ with poles of order at most
$l=(l_j)_{j=1}^N$ and $m=(m_j)_{j=1}^N$ at the punctures. The
universal configuration space $\M_g(l,m)$ can then be defined
as a moduli space of $\C$ under certain constraints on the set
of algebraic geometrical data \cite{kr96,krp96}. In this case
the space $\M_g(l,m)$ is a complex manifold with only orbifold
singularities. To introduce the local coordinates on
$\M_g(l,m)$ we cut apart the Riemann surface $\C$ along a
homology basis $A_i,~B_j$ $j=1,\ldots,g$ with canonical
intersection matrix
\bq
A_i\circ A_j=B_i\circ B_j=0\,,\qquad A_i\circ B_j=\d_{ij}\,.
\ll{cyrc}
\eq
By selecting cuts from $P_1$ to other $P_j$ for each $2\leq
j\geq N$ one gets a well-defined branch of the Abelian integrals
$y$ and $\l$.  Among the complete set of local coordinates on
$\M_g(l,m)$ the following moduli are distinguished
\bq
s_j=\oint_{A_j}yd\l\,,\qquad j=1,\ldots,g.
\ll{amod}
\eq
The universal configuration space $\M_g(l,m)$ is a base space
for a hierarchy of fibrations $\C^{(k)}(l,m)$ of particular
interest to us. These are the fibrations whose fiber above each
point of $\M_g(l,m)$ is the $k$-th symmetric power $S^k(\C)$ of
$\C$. This fiber $\C^{(k)}(l,m)$ is the set of all effective
divisors $D=\g_1+\cdots+\g_k$ (the $\g_j$'s may not be mutually
distinct) of ${\rm deg}\,k$ of $\C$, i.e. $\C^{(k)}(l,m)$
can be identified with the set of all unordered $k$-tuples
$\{\g_1,\ldots,\g_k\}$, where $\g_j$'s are arbitrary elements of
$\C$.

Let $\DD$ be the open set in $\M_g(l,m)$, where the zero
divisors of $dy$ and $d\l$ do not intersect. Fixing all the
local coordinates on $\M_g(l,m)$ except $s_j$ (\ref{amod}) one
can determine a smooth $g$-dimensional foliation of $\DD$,
independent of the choice we made to define the coordinates
themselves \cite{krp96}.  Hereafter, by abuse of notation, one
leaf of this foliation, is denoted just by $\M$ and $\C^{(k)}$
means the above fibrations restricted to $\M$.

Let $dS= yd\l$ be a meromorphic $1$-form on $\C$ with
the special Abelian integrals $y$ and $\l$, which
have fixed expansions near the punctures $P_j$ \cite{krp96}.
It means that we have imposed the certain constraints on the algebraic
geometrical data (according to \cite{kr96}
we used admissible data). These constraints ensure the
existence of global system of action-angle variables
and the presence of the corresponding symplectic form
\cite{neh72}.
The fact that we impose some constraints provides us with
additional properties of $dS$.  Namely, generating $1$-form
$dS$ possesses the property
\bq
\dfrac{\p dS}{\p s_j}=
\dfrac{\p yd\l}{\p s_j}=dw_j\,,\qquad j=1,\ldots,g\,,\ll{dds}
\eq
where $s_j$ are action coordinates (\ref{amod}) on the moduli space
$\M$ and differentials of the angle variables
$dw_j$ form some basis of holomorphic differentials
(normally, even if differential is holomorphic, its
moduli-derivative is not).  Moreover, form $dS$
give rise to differentials spanning a whole space  $\H_1(\C)$
of holomorphic differentials.  Hence, for any generic divisor
$D=\g_1+\cdots\g_g$ on $\C$ the standard $2$-form on $\C^{(g)}$
\bq \w_g=\d\left(\sum_{j=1}^g y(\g_j)d\l(\g_j)\right)=
\sum_{j=1}^g \d y(\g_j)\wedge d\l(\g_j)=
\sum_{j=1}^g ds_j\wedge dw_j\,,\ll{wg}
\eq
is a desired holomorphic symplectic form $\w_g$ on $\C^{(g)}$.
The set of variables $\{s_j,w_j\}_{j=1}^g$ are the complete set
of action-angle variables on $J(\C)$.
These action-angle variables $\{s_j,w_j\}$ have been obtained
by generalizing the definition of actions introduced for
integrable systems on tori in the form of periods of holomorphic
differentials $dw_j$ along the elements of a homology basis in
\cite{arn89,fr87}.

Now we turn to the uniform St\"{a}ckel systems.
The corresponding algebraic curve (\ref{sthc}) is a
hyperelliptic curve given by an equation of the form
\bq
\C:\qquad y^2=\prod_{i=1}^{2g+1} (\l-\l_i)\,,\ll{c0}
\eq
and puncture $P$ is the point at infinity $\l=\infty$.
Recall that the moduli $\l_j$ of $\C$ are integrals of motion (\ref{sthc}).
Solution to the inverse Jacobi problem and associated
Abel-Jacobi map on $\C$ relate a set of the action-angle variables
and the separated variables.

Variables of separation $q_j(t)$ give solution to the inverse
Jacobi problem (\ref{stinv}).  The associated Abel-Jacobi map
$\U:{\rm Div}(\C)\to J(\C)$ is restricted to Lagrangian
submanifold $\C^{(k)}$
\bq \U:\quad\C^{(k)}\to J(\C)\,.\ll{rajm}\eq
Note that whenever we discuss the Abel-Jacobi map, we shall
tacitly assume that a base point $\g_0$ (\ref{stinv}) on $\C$
has already been fixed in an appropriate position.

Suppose that point $D=\g_1+\cdots+\g_k$, $k\leq g$ belongs to
$\C^{(k)}$. The differential of the Abel-Jacobi map
(\ref{rajm}) at the point $D$ is a linear mapping from the
tangent space $T_D(\C^{(g)})$ of $\C^{(g)}$  at the point $D$
into the tangent space $T_{\U(D)}(J(\C))$ of $J(\C)$ at the
point $\U(D)$
\[\U_D^*:\quad
T_D(C^{(k)})\to T_{\U(D)}(J(\C))\,.\]
Now suppose that $D$ is a generic divisor, and $z_j$ is a local
coordinate on $\C$ near the point $\g_j$. Then
$(z_1,\ldots,z_k)$ yields a local coordinate system near the
point $D$ in $\C^{(k)}$. Let $dw_k$ ($k=1,\ldots,g$) is a basis
for a space $\H_1(\C)$ of holomorphic differentials on $\C$,
and near $\g_j$
\[dw_k=\phi_{kj}(z_j)dz_j\,,\]
where $\phi_{kj}(z_j)$ is holomorphic. It follows that the
Abel-Jacobi map $\U$ can be expressed near $D$ as
\[
\U(z_1,\ldots,z_k)=\left(
\sum_{j=1}^k\int_{\g_0}^{z_j} \phi_{1j}(z_j)dz_j,\ldots,
\sum_{j=1}^k\int_{\g_0}^{z_j} \phi_{gj}(z_j)dz_j\right)\,.\]
Hence
\bq
\U_D^*=
\left(\begin{array}{ccc}
\phi_{11}(\g_1)&\cdots&\phi_{g1}(\g_1)\\
\vdots&\ddots&\vdots\\
\phi_{1k}(\g_k)&\cdots&\phi_{gk}(\g_k) \end{array}\right)\,.\ll{bnm}
\eq
is the so-called Brill-Noether matrix.

\begin{th}
Transpose Brill-Noether matrix $\U^*_D$ on the genus $g\geq n$
hyperelliptic curve $\C$, which is the derivative of the
Abel-Jacobi map $\U$ at generic divisor $D\,, {\rm deg}D=n$, is
equal to the St\"ackel matrix $\bs$ for the uniform St\"ackel
system on $\bC^{2n}$ with metric \[{\rm
\gm}_{jj}(q_1,\ldots,q_n)=\dfrac{\det\bs}{\bs^{j1}}\,.\]
\end{th}
At generic point $D\,, {\rm deg}D=g$ matrix $\bs=U_D^{*t}$ is regular
matrix satisfying the St\"ackel theorem.

At $g>n$ we have to consider restriction of the Abel-Jacobi map
(\ref{rajm}) onto $\C^{(n)}$.  In this case symplectic form
$\w_n$ on $\C^{(n)}$ is an appropriate projection of $\w_g$
(\ref{wg}) and $\C^{(n)}$ be a Lagrangian submanifold in the
phase space $\bC^{2n}$.  The separated variables
$\{p_j,q_j\}_{j=1}^n$ are constructed from the first $2n$
action-angle variables (\ref{sepv}) only and the
action differential $dS=\sum_{j=1}^n p_jdq_j$ give rise to
an $n$-dimensional chart of the whole space $\H_1(\C)$.
The corresponding $n\times n$ St\"{a}ckel matrix is the left upper
$n\times n$ block of the general matrix $\bs=U_D^{*t}$
and, therefore, unless otherwise indicated, we assume
$n=g$.

As an example, let us consider some basis for
$\H_1(\C)$, for instance
\bq
dw_j=\dfrac{\l^{j-1}}{y(\l)}d\l\,,\qquad j=1,\ldots,g\,.\ll{bd0}
\eq
By choosing this basis we fix a basis of action-angle variables
(\ref{amod}-\ref{wg}).  To solve the Jacobi inversion problem
(\ref{stinv}) one gets variables of separation
\bq
p_j=y(\g_j)\,,\qquad
q_j=\l(\g_j)\,,\quad j=1,\ldots,g \ll{sepv}
\eq
for  a generic point $D=\g_1+\cdots+\g_g$ on $\C$, which
coincides with divisor of simple poles of the corresponding
Baker-Akhiezer function \cite{kr96}.  In the real case (when
$p_j$ and $q_j$ are real), the separated variables $q_j$
(\ref{sepv}) (so-called root variables) vary along cycles $A_j$
(\ref{cyrc}) over basic cuts on $\C$ and, therefore, our
problem is defined on $g$-dimensional real torus.
The holomorphic symplectic form $\w_g$ on $\C^{(g)}$ coincides with
standard ones (\ref{stw}) and a fiber $\C^{(g)}$ be a complex
Lagrangian submanifold of the phase space $\bC^{2g}$ (\ref{stl})
\bq
\C^{(n)}\equiv S^n(\C):\qquad
\left(\C(\l)\times\C(\m)\times\cdots\times\C(\nu)\right)/\s_n\,,
\qquad n\leq g\,,
\ll{slag}
\eq
where $\s_n$ is the permutation group on $n$ letters.

Recall, that derivative $\U_D^*$ bears a great resemblance to
the usual Gauss mapping. The map $\U^*_D$ induces a canonical
mapping from $\C$ into the $(g-1)$-dimensional projective space
$\C\to\bP^{g-1}$. On the other hand, the canonical mapping is
defined the derivative of the Abel-Jacobi map.  For a
hyperelliptic curve $\C$ of genus $g\geq 2$, the canonical map
$\C\to\bP^{g-1}$ is the composition of the double covering map
$\C\to\bP^1$, sending $(y,\l)$ to $\l$, with the Veronese map
$\bP^1\to\bP^{g-1}$ given by a basis for the polynomial ring of
degree $g-1$. With respect to the basis of $\H_1(\C)$
(\ref{bd0}), the canonical map of $\C$ has an extremely simple
expression \[(y,\l)\to
\l\to[\l^{g-1},\l^{g-2}\,\ldots,\l,1]\,.\] By using this map we
introduce the $g\times g$ matrix
\bq
\bs(\l,\m,\ldots,\nu)=
\left(\begin{array}{cccc}
\l^{g-1}&\m^{g-1}&\cdots&\nu^{g-1}\\
\vdots&\vdots&\ddots&\vdots\\
\l&\m&\cdots&\nu\\
1&1&\cdots&1 \end{array}\right)\,.\ll{stml}
\eq
determined on a Lagrangian submanifold (\ref{slag}).
For a generic point $D=\g_1+\cdots+\g_g$ in (\ref{sepv})
the St\"{a}ckel matrix is equal to
\bq
\bs(q_1,q_2,\ldots,q_g)=
\left.\bs(\l,\m,\ldots,\nu)
\right|_{\l=q_1,\m=q_2,\ldots,\nu=q_g}\,,\qquad
\sm_{kj}(q_j)=\left.\l^{g-k}\right|_{\l=q_j}\,.
\ll{stmd}
\eq
Recall, that the diagonal metric $\gm_{jj}$ is completely
determined by the corresponding St\"ackel matrix (\ref{stlam}).
Nevertheless, we introduce another equivalent definition of the
metric.  Substituting the St\"{a}ckel matrix (\ref{stmd}) in
the algebraic equation (\ref{stm}) one gets
\ben
&&\sum_{j=1}^g \sm_{kj}(q_j) \gm_{jj}(q_1,q_2,\ldots,q_g)=\d_{k1}=\nn\\
\ll{aleq}\\
&&
\sum_{j=1}^g\left.{\rm Res}\right|_{\l=q_j}\,
\dfrac{\l^{k-1}}{e(\l)}
=\dfrac{1}{2\pi i}\oint_C \dfrac{\l^{k-1}}{e(\l)}\,.\nn
\en
where by definition
\[
\gm_{jj}(q_1,q_2,\ldots,q_n)=
\left.{\rm Res}\right|_{\l=q_j}\,\dfrac{1}{e(\l)}\,,
\]
Here we introduced function $e(\l)$, which has zeroes at the
points $q_j$ giving solution of the inverse Jacobi problem.

In general, function $e(\l)$ with $g$ zeroes, which are solution
of inverse Jacobi problem, is expressed in the
Riemann theta-function
\bq
e(\l)=\theta\left(\U(\g_1,\ldots,\g_g)-\b-K\right)\,,
\qquad \U(\g_1,\ldots,\g_g)=\U(\g_1)+\cdots+\U(\g_g)\,.
\ll{gel}
\eq
Here $K$ is a vector of the Riemann constants and
$\b=(\b_1,\ldots,\b_g) \in \bC^g$ is a fixed
vector \cite{du81}. The principal properties of the function
$e(\l)$ (\ref{gel}) are considered in \cite{du81}.
\begin{prop}
Function $e(\l)$ on $\C$ with $g$ zeroes $(p_j,q_j)$ giving
solution to the Jacobi inversion problem is completely defined the
metric ${\rm \gm}_{jj}(q_1,q_2,\ldots,q_n)$ (\ref{aleq})
for a uniform St\"ackel system.
\end{prop}
We prove this proposition in the polynomial ring only.
In this case
\bq
e(\l)=\prod_{k=1}^g (\l-q_k)\,,\ll{el}
\eq
and
\bq
\gm_{jj}(q_1,q_2,\ldots,q_g)=
\left.{\rm Res}\right|_{\l=q_j} e^{-1}(\l)
=\dfrac{1}{\prod_{j\neq k}^g (q_j-q_k) }\,.
\ll{smetr}
\eq
To prove (\ref{aleq}) for this metric, it suffices to
consider the following integral
\bq
\dfrac{1}{2\pi i}\oint_C \dfrac{\l^k}{e(\l)}=
\sum_{j=1}^g \left.{\rm Res}\right|_{\l=q_j}\,\dfrac{\l^k}{e(\l)}
=-\left.{\rm Res}\right|_{\l=\infty}\,\dfrac{\l^k}{e(\l)}
=\d_{k,g-1}\ll{aint}
\eq
where $C$ encloses all $q_j$.

Function $e(\l)$ is defined on the universal
configuration space, i.e. it is independent on the moduli
$\l_j$ of $\C$ (integrals of motion) and on a choice of the basis  of
holomorphic differentials in $\H_1(\C)$.  For instance,  in the
polynomial ring let us consider a set of the equivalent
St\"{a}ckel matrices with the following entries (\ref{bnm})
\bq
\left.\sm_{kj}(\l)\right|_{q_j}=
\left.\phi_{kj}(\l)\right|_{q_j}\,,\qquad
\phi_{kj}(\l)=\l^{g-k}+a_1^{(j)}\l^{g-k-1}+
\ldots+a_{g-k}^{(j)}\,,\ll{sstm}
\eq
where polynomials $\phi_{kj}$ form various basises for the
polynomial ring of degree $g-1$. Substituting (\ref{sstm}) in
(\ref{aleq}) and (\ref{aint}) one obtains at once universal
solution $e(\l)$ (\ref{el}). Below we shall see that the
hamiltonian $H$ (\ref{sth}) with the diagonal metric
$\gm_{jj}$ (\ref{smetr}) is closely related to the distinguished puncture
$P$ at infinity $\l=\infty$ on the hyperelliptic curve $\C$
(\ref{c0}). The different St\"{a}ckel matrices (\ref{sstm})
correspond to the distinct sets of the integrals of motion
in the involution for a single hamiltonian $H$. The
completeness and functional independence of these integrals
directly follows from the completeness and independence of the
basis elements (\ref{sstm}) for a polynomial ring.

Finally, we look at other fibrations  $\C^{(n)}$ at $n\neq g$.
At $g>n$, to construct the metric $\gm_{jj}(q_1,\ldots,q_n)$
on $\C^{(n)}$,
we expand the initial curve $\C$ (\ref{c0}) by
\bq
y^2=\prod_{i=1}^{2g+1} (\l-\l_i)=
U_{2g+1}(\l)+\prod_{i=1}^{2n+1} (\l-\te{\l}_i)\,,
\qquad n\leq g\,.\ll{cp}
\eq
Here $U_{2g+1}(\l)$ is an at most $2g+1$ order
polynomial, which is regarded as a potential in
(\ref{sthc}). The  $n\times n$ St\"{a}ckel matrix
and the corresponding function $e(\l)$ may be associated to
the auxiliary genus $n$ curve
\bq
\te{\C}:\qquad \te{y}^2=
\prod_{i=1}^{2n+1} (\l-\te{\l}_i)\,.\ll{auc}\eq
Function $e(\l)$ are independent on the moduli of $\C$ (\ref{cp})
and, therefore, uniform potential $U_{2g+1}$ in (\ref{cp}) has
an arbitrary form and decomposition (\ref{cp}) determines the
highest power of the polynomial $U(\l)$ only.

At $n>g$ the above holomorphic symplectic form $\w_g$ on the leaves $\M$
is degenerate. However, a non-degenerate form on $\C^{(n)}$ may
be obtained by restricting $\C^{(n)}$ to the larger leaves
$\te{\M}$ of the foliation \cite{krp96}.  The leaves $\te\M$
correspond to the level sets of all the local coordinates
except to holomorphic $s_j$ (\ref{amod}) and to some additional
$(n-g)$ coordinates associated to meromorphic differentials
$d\te{w}_j$ in (\ref{dds}-\ref{wg}).  In fact, to construct
the action-angle variables we have to add
several meromorphic differentials to holomorphic angle
variables. Thus, at $n>g$ the symplectic $2$-form $\w_n$ on
$\C^{(n)}$ is meromorphic \cite{krp96}.

As an example, at $n=g+1$, we can add one local coordinate
in the neighborhood of puncture $P$ at infinity \cite{krp96}.
This additional coordinate occurs in the St\"{a}ckel matrix and in
the metric in the following way
\ben
&&\bs^{(g+1)}(\l,\m,\ldots,\nu)=q_0\bs^{(g)}(\l,\m,\ldots,\nu)\,,\nn\\
\ll{0metr}\\
&&e(\l)=q_0\prod_{j=1}^n(\l-q_j)\qquad
\gm_{00}=\left.{\rm Res}\right|_{\l=\infty}
\dfrac{\l^{g-1}}{\gm(\l)}\,.\nn
\en
At $n>g$ these systems with meromorphic form $\w_n$ possess
several reductions of the additional meromorphic coordinates,
for instance $q_0=const$ in (\ref{0metr}) \cite{kuz92}.

Above formulas are well adjusted for generalization.  If the
curve $\C$ (\ref{c0}) is substituted by
\bq \C:\qquad y^2=F(\l)=\dfrac{P_l(\l)}{Q_m(\l)}=
\dfrac{ \prod_{j=1}^{2g+1}(\l-\l_j)}{\prod_{k=1}^m(\l-\d_k)}\,,
\qquad m\leq 2g+1\,,
\ll{c1}
\eq
where $\{\d_k\}$ is a set of $m$ arbitrary constant,
one gets
\bq
\left.\sm_{kj}(\l)\right|_{\l=q_j}=
\left.\dfrac{\phi_{kj}(\l)}{Q_m(\l)}\right|_{q_j}\,,\qquad
e(\l)=\dfrac{\prod_{j=1}^g(\l-q_j)}{Q_m(\l)}\,.
\ll{metrq}
\eq
Note, that the algebraic equation (\ref{aleq}) is covariant
with respect to the transformations
\[\bs\to R^{-1}(\l)\bs\,,\qquad e(\l)\to R^{-1}(\l)e(\l)\,,\]
that leads to the general form of the metric
\bq
\gm_{jj}(q_1,q_2,\ldots,q_n)=\left.{\rm Res}\right|_{\l=q_j}\left(
\dfrac{Q_m(\l)R(\l)}{\prod_{j=1}^g(\l-q_j)}\right)\,.
\ll{genmetr}
\eq
associated to the curve $\C$.  We shall use this
freedom to consider the standard curvilinear coordinate systems
\cite{eis34,vil68,ka86}.

So, the hyperelliptic genus $g$ curve $\C$ may be associated to
a family of the uniform St\"{a}ckel systems on the phase
space $\bC^{2n}$ by using the Abel-Jacobi map $\U$, its
differential $\U^*_D$ and their restrictions on $\C^{(n)}$.
Diagonal metric $\gm_{jj}(q_1,q_2,\ldots,q_n)$
(\ref{stlam}) in the hamiltonian (\ref{sth}) is completely
defined by number of degrees of freedom $n$ and potential
$U(\l)$ is at most $2g+1$ order arbitrary polynomial.

On the other hand, one fixed metric $\gm_{jj}(q_1,\ldots,q_n)$
may be associated to an infinite set of the
hyperelliptic curves $\C$.  The corresponding hamiltonian
systems differ from each other by the power and by the form of
polynomial potentials $U(\l)$ (\ref{cp}).  Among these systems
we must to distinguish systems on $\C^{(n)}$ at $n>g$
(\ref{slag}) for which the number of degrees of freedom $n$ is
more than genus $g$ of the associated curve $\C$. In this case
the corresponding symplectic $2$-form on $\C$ is meromorphic \cite{krp96}.
In the next section, we shall identify these systems with the
degenerate or superintegrable systems \cite{ts97a}. Recall, that
for degenerate system the number of independent integrals of motion
is more than number degrees of freedom.


\section{The Lax representations}
\setcounter{equation}{0}
Let us recall that the key idea, which has started the modern
age in the study of classical integrable systems, is to bring
them into the Lax form. All the properties of the uniform
St\"ackel systems may be recovered from the properties of the
Abel map. Nevertheless, now we want to obtain the Lax
representations for all the uniform St\"{a}ckel systems
associated to the hyperelliptic curve $\C(y,\l)$ (\ref{c0}).
We consider construction of the Lax representation as
a necessary intermediate step to study quantum counterparts of
the St\"ackel systems.

In the simplest case  the Lax matrices $L(\l)$ or $L(y)$ are
defined as the matrix valued functions on  bare spectral curves
$F_\l$, $\l\in F_\l$ (\ref{bac}) or  $F_y$, $y\in F_y$, while
the full spectral curve $\C(y,\l)$ is given by the Lax
eigenvalue equations
\bq\C:\qquad
\det\left(L(\l)-y\right)=0\,,\qquad
\det\left(L(y)-\l\right)=0\,.\ll{fulc}\eq
As a result, $\C$ arises as a ramified covering over
the bare spectral curve $F_\l$ or $F_y$ \cite{kr80}.

Till now a delicate questions is how to construct the Lax
matrices $L(\l)$ or $L(y)$ for a given integrable system. The
one integrable system may be associated to the different curves
and one curve $\C$ may be associated to the
different mechanical integrable system on a common phase space.
As an example, the $n$-particles Toda lattice can be
equivalently formulated in terms of two different Lax
representations \cite{ft87} associated to the single
hyperelliptic curve $\C$.

Here we consider equation for a general algebro-geometric
symplectic structure associated to the spectral curve $\C$ of
the given Lax representation $L$
\bq
\w_n=-\sum_\a {\rm Res}_{P_\a}
\dfrac{<\d\psi^+\wedge \d L \psi>}{<\psi^+\psi>}\ll{laxeq}
\eq
proposed in \cite{krp96}. Here $\w_n$ is the restriction of the
algebro-geometrical symplectic form (\ref{wg}) on $\C$
generated by two differentials $dy$ and $d\l$ having poles at
punctures $P_\a$.  Functions $\psi$ and $\psi^+$ are the
Baker-Akhiezer function on $\C$ and it's dual function.  If we
fix some $2$-form $\w_n$ and the Baker-Akhiezer functions
$\psi$, $\psi^+$ on  a given curve $\C$, then one can attempt
to recover the associated Lax matrix $L$.

For the some particular St\"ackel systems the $2\times
2$ Lax matrices \cite{kuz92,mum8412} and the corresponding vector
Baker-Akhiezer function $\vec\psi$ associated to natural
vector fields on the Jacobian of any hyperelliptic curve are known.
On the other hand, we know the general scalar Baker-Akhiezer
function $\psi$ on $C$ defined by its analitical properties on
$\C$, which corresponds to geodesic systems with diagonal
metric \cite{kr96}.

Note, here we have the vector Baker-Akhiezer function
$\vec\psi$, which is the eigenfunction of the matrix $L$
associated to the curve $\C$, and scalar Baker-Akhiezer $\psi$,
which is completely defined by analitical properties on the
same curve $\C$.

For the uniform St\"ackel systems
let us identify the preassigned symplectic structure $\w$
(\ref{stw}) with the symplectic structure (\ref{wg})
defined on a hyperelliptic algebraic curve $\C$.
Next we try to recover Lam matrix for a geodesic motion
under the following additional assumptions:
{\it
\par 1) $L(\l)$ is a generic $2\times 2$ matrix associated
to a spectral hyperelliptic curve $\C$ of genus $g=[(n-1)/2]$.
\par 2) The associated vector Baker-Akhiezer function $\vec{\psi}$
has a constant normalization $\vec{\a}$ \cite{skl95}
\[<\vec{\a},\vec{\psi}>=\a_1\psi_1+\a_2\psi_2=1\,,\qquad
\vec{\a}=(0,1)\,.\]
\par 3) The first component of $\vec{\psi}$ in (\ref{laxeq})
is proportional to the unique Baker-Akhiezer function $\psi$
on $\C$ with fixed analytical properties \cite{kr96}.}

At the first assumption $n$ is a number of integrals of motion,
which are moduli of $C$ $(n=2g+1)$ and, therefore, form $\w_n$
in (\ref{laxeq}) is a restriction of meromorphic symplectic
form $\w_g$ (\ref{wg}) to  the minimal $n$-dimensional leaf $\te{\M}$
\cite{krp96} for integrable systems on $\bC^{2n}$.
The second assumption allows us to reduce
vector Baker-Akhiezer function to scalar one.
In this case, solution of (\ref{laxeq}) is completely defined
by the function $\psi$ on $\C$ only.
At first we present this particular
solution in term of the function
$e(\l)$ associated to the Abel map $\U$.
Introduce function $e(\l)$ and its time derivative
\bq
e(\l)=\dfrac{\prod_{j=1}^n(\l-q_j)}{\prod_{k=1}^m(\l-\d_k)}
\,,\quad m\leq n\,,\qquad e_x(\l)=\{H,e(\l)\}\,,\ll{geom}
\eq
where $\{\d_k\}$ is a set of $m$ arbitrary constant and
$H$ be a hamiltonian of the geodesic motion
\bq
H=\sum_{j=1}^n \gm_{jj}(q_1,\ldots,q_n)p_j^2\,,
\qquad\gm_{jj}(q_1,\ldots,q_n)=
\left.{\rm Res}\,\right|_{\l=q_j}\dfrac{1}{e(\l)}\,.\ll{geo}
\eq
Thus, in the Lax equation for a geodesic motion
\[L_x(\l)=\{H,L\}=\left[L,A\right]\,,\]
matrices $L$ and $A$ are given by
\ben
L(\l)&=&\left(\begin{array}{cc}-{e_x}/2&e\\
-e_{xx}/2&{e_x}/2\end{array}\right)(\l)\equiv
\left(\begin{array}{cc}h&e\\f&-h\end{array}\right)(\l)\,,\nn\\
\ll{sew}\\
A(\l)&=&\left(\begin{array}{cc}0&1\\0&0\end{array}\right)\,.\nn
\en
The hamiltonian $H$ (\ref{geo}) is equal to a highest
residue at the distinguished Weierstrass point on $\C$ at
infinity $\l=\infty$
\bq H=-\left.{\rm
Res}\right|_{\l=\infty} \l^{n-m} {\rm det} L(\l)\,.\ll{hinf}
\eq
where the full spectral curve $\C$ is equal to
\ben
\C:\qquad &&y^2=F(\l)\equiv\det L(\l)\,,\nn\\
\ll{dec1}\\\
&&F(\l)=-h^2(\l)-e(\l)f(\l)=
\dfrac{e\cdot e_{xx}}{2}-\dfrac{e_x^2}{4}\,.\nn
\en
By definition  zeroes of $e(\l)$ be separation variables
and conjugated variables $p_j$ are given by
\bq
p_j=\left.h(\l)\right|_{\l=q_j}\,,\qquad
h(\l)=-e_x/2=e(\l)\sum_{j=1}^n
\dfrac{\gm_{jj}(q_1,\ldots,q_n)p_j}{\l-q_j}\,.
\eq
In accordance with \cite{skl95}
pairs of separation variables $(q_j,p_j)$ lie on the
spectral curve $\C$
\[y^2(\g_j)=p_j^2=\left.h^2(\l)\right|_{\l=q_j}=-
F(\l=q_j)=-\left.F(\l)\right|_{\g_j}
\,.\]

As usual, rational function $F(\l)$  admits some different
representations
\bq
F(\l)=\dfrac{\sum_{j=1}^{n} I_j\l^{n-j}}{\prod_{k=1}^m (\l-\d_k)}=
\sum_{k=1}^m\dfrac{J_k}{(\l-\d_k)}+\sum_{k=m+1}^n
J_{k}\l^{n-k-1}\,.\ll{ddec}
\eq
Here $\{I_j\}_{j=1}^n$ and $\{J_k\}_{k=1}^n$ are two sets of
independent integrals of motion in the involution.  The first
set of integrals $\{I_j\}_{j=1}^n$ in (\ref{ddec}) corresponds
to the St\"{a}ckel matrix (\ref{stml}).  The set of equivalent
St\"{a}ckel matrices (\ref{sstm}) relate to another
decompositions of a numerator of $F(\l)$ (\ref{ddec}). The
second set of integrals $\{J_k\}_{k=1}^n$ in (\ref{ddec}) is
associated to an expansion near punctures $\{\d_k,\infty\}$ on
$\C$.

The spectral curve $\C$ (\ref{fulc}) is a time-independent
curve and, therefore,
\bq
\{H,F(\l)\}=0\,,\qquad\Rightarrow\qquad
\p^3_x e(\l)=e_{xxx}=0\,.\ll{e3}
\eq
Thus, in fact \cite{cf97}, we consider the polynomial solutions
$e(\l)=\prod (\l-q_j(t))$ to the equation (\ref{e3}) and
describe the hamiltonian dynamics of their zeroes $q_j(t)$
(recall, that $\p_x$ means derivative by time).

Substituting function $e(\l)$ (\ref{geom}) and
hamiltonian $H$ (\ref{geo}) into (\ref{e3}) one gets the
equations in the metric $\gm_{jj}=\HH^2_j$ (\ref{metr}).
If we introduce  so-called rotation coefficients
\bq
\b_{ij}={\p_i \HH_j\over \HH_i}, \ \ i\neq j, \label{rotcf}
\eq
these equations may be reduced to the following equations
\cite{zah96}
\ben
&&\p_k \b_{ij}=\b_{ik}\b_{kj},\ \ i\neq j\neq k, \nn\\
\ll{lame1}\\
&&\p_i \b_{ij}+\p_j\b_{ji}+\sum_{m\neq i, j} \b_{mi}\b_{mj}=0,
\ \ i\neq j,\nn
\en
where the notation $i\neq j\neq k$ means that indices $i,j,k$
are distinct.

Of course, these equations may be obtained without any Lax
representation by using definition (\ref{stlam}) of the metric,
properties of the Abel-Jacobi map and preassigned asymptotic
behavior of $e(\l)$ at the distinguished point $\l=\infty$.

The equations (\ref{lame1}) are equivalent to
the vanishing conditions of all a'priory non-trivial components
of the curvature tensor \cite{zah96,kr96,razs96}.  Therefore,
using (\ref{lame1}) we conclude that local Riemannian
submanifold $({\cal V}_n,\left.\gm\right|_{\cal V})$
(\ref{hjeq}) of the Riemannian manifold $(\bC^n,\gm)$ is a flat
manifold whose metric is diagonal with respect to the
coordinates $\{q_j\}$.  Imposing some additional restrictions
on the space of solutions to (\ref{laxeq})\cite{kuz92},
one could get the Bourlet type equations
\cite{razs96} related to another Riemannian manifolds of constant
curvature.

To construct more general solutions to (\ref{laxeq}) associated
to hyperelliptic curve $\C$ of higher genus  we begin with
calculation of the Poisson bracket relations for the initial
Lax matrix $L(\l)$.  It allows us to identify the space of solutions
to equation (\ref{laxeq}) with the loop algebra $\L(sl(2))$ in fundamental
representation \cite{ft87} and then to use the representation
theory of the underlying algebra $sl(2)$ \cite{ts96b}.
\begin{th}
The Poisson bracket relations  for the matrix $L(\l)$ (\ref{sew})
are closed into the following $r$-matrix algebra at $m\leq n$ only
\bq
\{{\on{L}{1}}(\l),{\on{L}{2}}(\m)\}=
[r_{12}(\l,\m),{\on{L}{1}}(\l)]-[r_{21}(\l,\m),{\on{L}{2}}(\m)\,]\,.
\ll{rpoi}
\eq
\end{th}
Here the standard notations are introduced:
\ben
&&{\on{L}{1}}(\l)= L(\l)\otimes I\,,\qquad
{\on{L}{2}}(\m)=I\otimes L(\m)\,,\nn\\
\ll{r}\\
&&r_{12}(\l,\m)=\dfrac{\Pi}{\l-\m}\,\qquad
r_{21}(\l,\m)=\Pi r_{12}(\m,\l)\Pi\,,\nn
\en
and $\Pi$ is the permutation operator of auxiliary spaces \cite{ft87}.

The Poisson bracket relations for
the Lax matrix $L(\l)$ (\ref{sew}) are preassigned by the
initial symplectic structure (\ref{wg}).  It is necessary to
calculate two brackets only
\bq
\{e(\l),e(\m)\}=0\,,\ll{br1}
\eq
and
\ben
\{h(\l),e(\m)\}&=&
\left\{ e(\l)\sum_{j=1}^n \dfrac{\gm_{jj}(q_1,\ldots,q_n)\,p_j}{\l-q_j},
\dfrac{\prod_{j=1}^n(\l-q_j)}{\prod_{k=1}^m(\l-\d_k)}\right\}\nn\\
&=&
-e(\l)e(\m)\sum_{j=1}^n\dfrac{\gm_{jj}}{(\l-q_j)(\m-q_j)}\nn\\
\ll{br2}\\
&=&\dfrac{e(\l)e(\m)}{\l-\m}
\sum_{j=1}^n \left(\dfrac{\gm_{jj}}{\l-q_j}-
\dfrac{\gm_{jj}}{\m-q_j}\right)
=\dfrac{1}{\l-\m}\left[e(\m)-e(\l)\right]\,,\nn
\en
where we used a standard decomposition of rational
function
\[e^{-1}(\l)=
\sum_{j=1}^n\dfrac{\gm_{jj}}{\l-q_j}\,,\qquad
\gm_{jj}=\left.{\rm Res}\right|_{\l=q_j} e^{-1}(\l)\,.\]
Another Poisson brackets may be directly derived from these
brackets and by definition of the entries of the Lax matrix $L(\l)$
(\ref{sew}) via derivative of the single function $e(\l)$
\ben
\{h(\l),h(\m)\}&=&0\,,\nn\\
\{f(\l),e(\m)\}&=&\p_x
\{h(\l),e(\m)\}=\dfrac{2}{\l-\m}\left[h(\l)-h(\m)\right]\,,\nn\\
\ll{br3}\\
\{f(\l),h(\m)\}&=&-\dfrac12\p^2_x
\{h(\l),e(\m)\}=\dfrac{1}{\l-\m}\left[f(\l)-f(\m)\right]\,,\nn\\
\{f(\l),f(\m)\}&=&-\dfrac12\p^3_x\{h(\l),e(\m)\}=0\,,\nn
\en
To derive the first bracket we have to combine second and first
derivatives of the brackets (\ref{br1}) and (\ref{br2}),
respectively. At the last bracket one substitutes the equation of
motion (\ref{e3}).

If, contrary to our geometric conventions,
the order of polynomial $Q_m(\l)$ is more then order of polynomial
$P_l$ in (\ref{c1}), i.e. if $m>n$ in the metric (\ref{sew}), then
rational function $e(\l)$ admits another representation
\[e^{-1}(\l)=
\sum_{j=1}^n\dfrac{\gm_{jj}}{\l-q_j}+\xi(\l,q_1,\ldots,q_n)\,,\]
where remainder $\xi(\l)$ is a certain polynomial.  Substituting
this function $e(\l)$ into (\ref{br1}-\ref{br3}) one gets
\[\dfrac{\p\xi(\l,q_1,\ldots,q_n)}{\p \l}=0\,.\]
This constraint to remainder $\xi(\l,q_1,\ldots,q_n)$ directly
follows from the symmetry of the last Poisson bracket in
(\ref{br3}).

The $r$-matrix algebra (\ref{rpoi}) is so-called linear case of
the $r$-matrix algebras corresponds to integrable systems,
which are modelled on coadjoint orbits of Lie algebra $sl(2)$.
The $r$-matrix in (\ref{r}) is a standard rational $r$-matrix
on $\L(sl(2))$ \cite{rs87}.  The general form of the function
$e(\l)$ (\ref{gel}) leads to the elliptic and trigonometric
$r$-matrices \cite{rs87,krw95}

Thus, for a geodesic motion (\ref{geo}) the Lax representation
(\ref{sew}) with  arbitrary poles $\{\d_k\}_{k=1}^m$
(\ref{geom}) may be regarded as a generic point at the loop
algebra $\L(sl(2))$ in fundamental representation after an
appropriate completion \cite{rs87}.  Since, to construct the
Lax representation for a potential motion with the fixed metric
$\gm_{jj}(q_1,\ldots,q_n)$ (\ref{geom}) we can use the
outer automorphism of the space of infinite-dimensional
representations of $sl(2)$ proposed in \cite{ts96b}.

Applying this automorphism of the
underlying algebra $sl(2)$ directly to the Lax representation
$L(\l)$ (\ref{sew}) on $\L(sl(2))$
we obtain a family of the new Lax pairs
\ben
&&L'(\l)=L(\l)-\s_-\cdot\left[\phi(\l)e^{-1}(\l)\right]_{N}\,,\qquad
\s_-=\left(\begin{array}{cc}0&0\\
1&0\end{array}\right)\,,\nn\\
\ll{mapmn}\\
&&A'(\l)=A-\s_-\cdot\left[\phi(\l)e^{-2}(\l)\right]_{N}=
\left(\begin{array}{cc}0&1\\
u_{N}(\l)&0\end{array}\right)\,.\nn
\en
Here $\phi(\l)$ is a function on spectral parameter and
$[z]_{N}$ means restriction of $z$ onto the ${\rm
ad}^*_R$-invariant Poisson subspace of the initial $r$-bracket
\cite{ts96b,krw95,eekt94}.  For the rational $r$-matrix (\ref{r})
we can use the linear combinations of the following Taylor
projections
\bq
{[ z ]_{N}}=\left[\sum_{k=-\infty}^{+\infty} z_k\l^k\,
\right]_{N}\equiv
\sum_{k=0}^{N} z_k\l^k\,,
\ll{cutmn}
\eq
or the Laurent projections \cite{ts96b,krw95}.

The mappings (\ref{mapmn}) from the representation of the
loop algebra $\L(sl(2))$ to representations of
the universal enveloping algebra $U(\L)$
play the role of a dressing
procedure allowing to construct the Lax matrices $L'_{N}(\l)$
for an infinite set of new integrable systems starting from the
single known Lax matrix $L(\l)$ associated to one integrable
model.  This mapping preserves the metric
$\gm_j(q_1,\ldots,q_n)$ in the hamiltonian (\ref{sth}), but
changes the potential $U(q_j)$ and associated curve $\C$.

New Lax matrix $L'(\l)$ (\ref{mapmn}) obeys the linear
$r$-bracket (\ref{rpoi}), where constant $r_{ij}$-matrices
substituted by $r_{ij}'$-matrices depending on dynamical
variables \cite{ts96b,krw95}.
\bq r_{12}(\l,\mu)\to
r'_{12}=r_{12}
-\frac{\left( [\phi(\l) e^{-2}(\l)]_{N}-
[\phi(\m) e^{-2}(\m)]_{N}\,\right)}{(\l-\m)}\cdot\s_-\otimes\s_-
\,.\ll{dpr}
\eq

We have to distinguish systems on $\C^{(n)}$ at $n>g$ (\ref{slag}) for
which the number of degrees of freedom $n$ is more than genus
$g$ of the associated curve $\C$. According to \cite{krp96} the
corresponding symplectic form is meromorphic.
In this case the action differential $dS=yd\l$ give rise to a
whole space $\H_1(\C)$ and, in addition, several meromorphic
differentials on $\C$.  We can identify these systems with
the degenerate or superintegrable systems \cite{ts97a}.
\begin{th}
The complete set of noncommutative integrals of motion for the
degenerate uniform St\"ackel systems with meromorphic
symplectic form $\w_g$ is determined by the generalized
spectral surface
\[
\C(y,\l,\m):\qquad
\det\left(yI+\Pi L'(\l)\otimes L'(\m)\right)=0\,.
\]
\end{th}
Here we used the outer product of the $2\times 2$ Lax
matrices $L'(\l)$ with $L'(\m)$ and $\Pi$ means $4\times 4$
permutation matrix in $\bC^2\times\bC^2$.
Equation of motion for the matrix $L(\l,\m)=
\Pi L'(\l)\otimes L'(\m)$ is equal to
\ben
\dfrac{d}{dt}L(\l,\m)&=&
L(\l,\m)A(\l,\m)-\Pi A(\l,\m)\Pi^{-1}L(\l,\m)\,,\nn\\
\ll{lleq}\\
A(\l,\m)&=&A(\l)\otimes I+I\otimes A(\m)\,,\nn
\en
where matrix $A(\l)$ is a second Lax matrix and $I$ is a unit matrix.

It is easy to derive from (\ref{mapmn}), that $n>g$ iff $n\geq
N$, where $N$ is a highest power in the Taylor projection
(\ref{cutmn}).  In this case the corresponding $r$-matrix
(\ref{dpr}) preserves the simple pole at the puncture $P$ at
$\l=\infty$ and the associated second Lax matrix $A'$ remains a
constant in spectral sense $\dfrac{\partial A(\l)}{\partial
\l}=0$ under the mapping (\ref{mapmn}).

Thus, for the degenerate systems $A(\l,\m)=\Pi
A(\l,\m)\Pi^{-1}$ and equation (\ref{lleq}) takes the standard
Lax form and it proves the theorem.

As usual, spectral curve $\C$ (\ref{fulc}) of $L'(\l)$ is a
generating function of the involutive family of integrals of
motion.  Substituting functions
$\phi(\l)=\l^{n}Q_m^{-1}(\l)U_N(\l)$ into $L'(\l)$
(\ref{mapmn}) one gets their spectral curve in the form
\[\C:\qquad y^2=F'(\l)=\det L'(\l)=U_N(\l)+
\dfrac{\sum_{j=1}^{n} I_j'\l^{n-j}}{\prod_{k=1}^m (\l-\d_k)}\,,\]
where $\{I'_j\}$ are integrals of motion.
It is a time-independent curve and, therefore,
\bq
\dfrac{d F'(\l)}{dt}=0\,,\quad\Rightarrow\quad
\left[\dfrac14\partial^3_x +u_{N}(\l)\partial_x
+\dfrac12 u_{N,x}(\l)\,\right]\cdot e(\l)=0\,.
\ll{sgeq}
\eq
Of course, this equation may be obtained directly
in framework of symplectic geometry \cite{er89}).
Let us briefly explain an origin of this equation in
the theory of nonlinear equation, that allows us to
relate scalar Baker-Akhiezer function $\psi$
and function $e(\l)$.

The same algebro-geometrical
symplectic form $\w_g$ (\ref{wg}) on hyperelliptic curve
$\C$ (\ref{c0}) leads directly to
a hamiltonian structure for soliton equations
\cite{nov82,krp96}.  As an example, we consider the KdV
equation associated to hyperelliptic curve  (\ref{c0}) with one
puncture $P$ ($N=1$) at infinity $\l=\infty$ and at $l=1$,
$m=2$ \cite{krp96}.  Let us select one leaf of foliation
corresponded to $d\l$ with all zero periods
\[\oint_C d\l=0\]
for an arbitrary cycle $C$. In this case, the Abelian integral
$\l(P)$ is a single-valued function, with only a pole of second
order at $P$ ($m=2$).
For finite-gap solutions of the KdV equations, moduli $s_j$
(\ref{amod}) are canonically conjugated with respect to the
Gardner-Faddeev-Zakharov symplectic structure to angle
variables $w_j$ (see \cite{fm76} and references within).
Thus, the uniform St\"ackel systems have a common set of the
action-angle variables with solutions of the KdV equations.

Starting with this  set of variables we
consider general algebro-geometric equation (\ref{laxeq})
for nonlinear systems.
Solution of the equation (\ref{laxeq}) in a ring of second order
differential operators with the standard Baker-Akhiezer function
$\psi$ on $\C$ (\ref{c0}) is well known \cite{nov82,du81,krp96,almar92}.
The associated Shr\"{o}dinger operator has the form
\bq
\L(\l)=-\dfrac{\p^2}{\p x^2}+u(x,t,\l)\,,\ll{shrd}
\eq
where $\l$ is a parameter. In some simple cases, such as the
KdV equation, this parameter $\l$ appears as an eigenvalue and one
ultimately equates the potential $u$ with a solution of the
nonlinear equation itself. Let us look for a solution $\A(\l)$
of the Lax system in the ring of differential operators
\ben
&&\L(\l)\psi=0\,,\nn\\
\ll{diflax}\\
&&
\left(\dfrac{\p\L(\l)}{\p t}+[\L,\A]\,\right)\,\psi=0\nn
\en
of the form
\bq
\A(\l)=e(\l)\dfrac{\p}{\p x}-\dfrac12\dfrac{\p e(\l)}{\p x}\,,
\ll{difa}
\eq
Substituting the given form of $\A$ into the Lax system,
one gets
\bq
\dfrac{\p u}{\p t}=
-2\left[\dfrac14\partial^3_x +u(\l)\partial_x
+\dfrac12 u_{x}(\l)\,\right]\cdot e(\l)\,.
\ll{geq}
\eq
Equation (\ref{geq}) is called the generating equation.  For a
different choices of the form of $e(\l)$ and $u(\l)$, this
procedure leads to different hierarchies of integrable
equations, as an example to the KdV, nonlinear Shr\"{o}dinger
and sine-Gordon hierarchies \cite{al81,almar92} or to the Dym
hierarchy \cite{alchm94}. If we consider the solutions of the
equation (\ref{geq}) in the form of polynomial (\ref{el}), then
the roots $q_j$ of $e(\l)$ define the root variables and as a
result finite-gap solutions of the problem of geodesic (see
\cite{al81,almar92,alchm94} and references within).

Substitution of the
special form of second operator $\A(\l)$ (\ref{difa}) into the
Lax system (\ref{diflax}) allows us to eliminate the
Baker-Akhiezer function $\psi$ and to construct $2\times2$ Lax matrix
in $e(\l)$.  In fact, we replace the
Baker-Akhiezer function $\psi$ on $\C$ to the mutually disjoint
function $e(\l)$ on $\C$, which has a transparent mechanical
interpretation (\ref{el}).  Recall, that function $e(\l)$ is
defined as function with zeroes, which give solution to the
Jacobi inversion problem \cite{du81} on the hyperelliptic curve
$\C$.


\section{The flat coordinates}
\setcounter{equation}{0}

According to \cite{kr96} at $n=g$ the orthogonal curvilinear coordinates
$\{p_j,q_j\}_{j=1}^g$ form a generic divisor of
the simple poles of the Baker-Akhiezer function $\psi$, which
is defined by their analytical properties on $\C$. The
evaluation of $\psi$ at a set of punctures on $\C$ determines
the flat coordinates $\{\sp_j,\sx_j\}_{j=1}^g$ for the diagonal
metric (\ref{metr}).  It turns out that up to constant factors
the Lam\'e coefficients $\HH_j$ are equal to the leading terms of
the expansion of the same function $\psi$ at the punctures on
$\C$ \cite{kr96}.

Next we reach the same conclusions by using the function $e(\l)$
and the corresponding Lax representation $L(\l)$ on $\C$.
 As usual, we reduce the study of algebraic
geometrical data to the analysis of the associated geodesic
motion. The crucial observation is that the equations of
motion in coordinates
$\{\sp_j,\sx_j\}_{j=1}^n$ on the Riemannian manifolds of
constant curvature have a Newton form and the corresponding
hamiltonian has a natural form
\bq
\ddot{\sx}_j=\xi_j(\sx_1,\ldots,\sx_n)\,,\qquad
H=\sum a_{ij} \sp_i\sp_j+V(\sx_1,\ldots,\sx_n)\,,\quad a_{ij}\in\bC\,,
\eq
where $\xi_j(\sx_1,\ldots,\sx_n)$ and potential $V(\sx_1,\ldots,\sx_n)$
are functions on coordinates
only. Let us introduce new function $\B(\l)$
\bq
\B^2(\l)=e(\l)=\HH^{-2}(\l)\,,\ll{bfl}
\eq
which is "inverse" to the Lam\'e  coefficients $\HH_j$ (\ref{stlam}).
One immediately gets
\bq
F(\l)={\B}^3{\B}_{xx}\,,\qquad
F'(\l)={\B}^3{\B}_{xx}+{\B}^4
\left[\dfrac{\phi(\l)}{{\B}^4}\right]_{N}\,,\label{dsub}
\eq
These equations have the form of Newton's equations for the
function $\B$
\ben
{\B}_{xx}&=&F(\l){\B}^{-3}\,,\nn\\
\label{newton}\\
{\B}_{xx}&=&F'(\l){\B}^{-3}-{\B}
\left[\dfrac{\phi(\l)}{{\B}^4}\right]_{N}\,,\nn
\en
To expand function $\B(\l)$ at the Lourent set
\[{\B}=\sum_{j=0}^N \sx_{N-j}\, \l^j\]
it is easy to prove that coefficients $\sx_j$ obey the Newton
equation of motion (\ref{newton})
(see (\ref{e3}) and references within \cite{cf97}).  Here we
reinterpret the coefficients of the bare curves $F(\l)$ and
$F'(\l)$ in (\ref{newton}) not as functions on the phase space,
but rather as integration constants.  In  variables $\sx_j$
mapping (\ref{mapmn}) affects only on the potential
($\sx$-dependent) part of the integrals of motion $I_k$.  The
kinetic (momentum dependent) part of $I_k$ remains unchanged.
So, the dressing mapping (\ref{mapmn}) allows us to get over
from a free motion on $\bC^{2n}$ to a potential motion on
$\bC^{2n}$.

As an example, from (\ref{newton}) we get some well known
orthogonal curvilinear coordinates on $\bR^n$
(see \cite{eis34,vil68,ka86}):
\begin{center}
elliptic coordinates $m=n$ in (\ref{metrq})
\[
e(\l,q_1,\ldots,q_n)=\dfrac{\prod_{j=1}^n (\l-q_j)}{\prod_{k=1}^n (\l-\d_k)}=
1+\sum_{k=1}^n\dfrac{\sx_k^2}{\l-\d_k}
=\B^2(\l,\sx_1,\ldots,\sx_n)\]
\[\d_1<\sx_1<\d_2\cdots<\d_n<\sx_n\]
parabolic coordinates $m=n-1$ in (\ref{metrq})
\[
e(\l,q_1,\ldots,q_n)=\dfrac{\prod_{j=1}^n (\l-q_j)}{\prod_{k=1}^{n-1}(\l-\d_k)}=
\l-\sx_n+\sum_{k=1}^{n-1}\dfrac{\sx_k^2}{\l-\d_k}
=\B^2(\l,\sx_1,\ldots,\sx_n)\]
\[\sx_1<\d_1<\sx_2\cdots<\d_{n-1}<\sx_n\]
spherical coordinates $m=n+1$ see (\ref{0metr})
\[
e(\l,q_0,\ldots,q_n)=\dfrac{q_0\prod_{j=1}^n (\l-q_j)}{\prod_{k=1}^{n+1}(\l-\d_k)}=
\sum_{k=1}^{n+1}\dfrac{\sx_k^2}{\l-\d_k}
=\B^2(\l,\sx_1,\ldots,\sx_{n+1})\]
\end{center}
Curvilinear coordinates $\{q_j\}$ are zeroes of function
$e(\l)$ and  flat coordinates $\{\sx_j\}$  are residues of
$e(\l)=\B^2(\l)$ at the punctures, in accordance with the
Baker-Akhiezer function approach \cite{kr96,skl95}.

All the separable orthogonal curvilinear coordinate systems in
$\bR^n$ may be obtained from these coordinate systems
\cite{ka86,vil68,kuz92}. According to \cite{kbm95}, all the
possible separable in these coordinates potentials, which are
polynomials or rational functions of the cartesian coordinates
$\sx_j$, belong to the set of the uniform St\"{a}ckel systems.
Thus, we can claim that every such mechanical system is embedded
into a proposed scheme.

\subsection{Quasi-point canonical transformations.}
In conclusion, we discuss another parameterizations of the
function $e(\l)$. Of course, function $e(\l)$ admits various
representations in different variables and we can use this
freedom, as an example to solve equations of motion
\cite{cf97}.  The considered above parametrization describe the
point canonical transformation only.  Here we discuss an
application of the Weierstrass reduction theory to construct
another cartesian coordinates on $\C$.

It is obvious, that the Lax representation
\[\dot{L}(\l)=\left[L(\l),A(\l)\right]\]
is covariant with respect to the transformation of the first
Lax matrix
\[L(\l)\to \phi(\l,\l_1,\ldots,\l_k)L(\l)\]
with an arbitrary function $\phi(\l,\l_1,\ldots,\l_k)$ on
time-independent moduli $\{\l_j\}$ of $\C$ and on spectral
parameter $\l$.  However, this transformation drastically
changes the Poisson bracket relations (\ref{r}) and
parameterization of $L(\l)$ in the flat coordinates
$\{\sp_j,\sx_j\}$. Hence, in addition to considered above flat
coordinates $\{\sp_j,\sx_j\}_{j=1}^n$, the same function
$e(\l,q_1,\ldots,q_n)$ may be associated to another set of flat
coordinates. Now we show that
to introduce these new variables $\{\sp_j,\sx_j\}$
we can use various covering of the initial curve $\C$, as an
example, covering listed in \cite{bak97}.

Let us assume that the initial torus $J(\C)=T^{2g}$ may be
decomposed in a direct product of several tori
\bq T^{2g}=T^{2g_1}\times\cdots\times T^{2g_k}\,,
\qquad \sum_{j=1}^k g_j=g\,.\ll{dt}\eq
The corresponding Riemann matrix has a block form
$B=B_{1}\times B_2\cdots\times B_k$,
where $B_j$ are the $g_j\times g_j$ Riemann matrices and the
corresponding Baker-Akhiezer function on $\C$ is factorized.
In this case we can consider curve $\C$ as a $K$-sheeted
covering of tori $T^{2g_j}$. Such covers are known to exist
for any $K>1$ and for arbitrary tori \cite{bak97}

First of all, we can introduce the separated variables
$\{q_j\}$ associated to a whole torus $T^{2g}$.
For dynamics on $J(\C)=T^{2g}$ the corresponding Lax
representations $L(\l)$ (\ref{sew}) are $2\times 2$ matrices.

Secondly, we can introduce another set of separated variables
$\{\te{q}_j\}$ associated to each torus $T^{2g_j}$ in (\ref{dt}).
For dynamics splitting on several tori $T^{2g_j}$ the Lax
representations have a block form
\bq
L(\l)=\left(\begin{array}{ccc}
L_{1}&~&~\\
~&\ddots&~\\
~&~&L_k\end{array}\right)(\l)\,,\ll{bllax}
\eq
where $L_j(\l)$ are the $2\times 2$ matrices defined by
functions $e_j(\l)$ on the each torus $T^{2g_j}$ \cite{krw95}.
Two sets of variables $\{q_j\}$ and $\{\te{q}_j\}$ are related
by canonical transformation induced by the covering, that allows us to
get $2\times 2$ Lax matrix instead of matrix (\ref{bllax}).
It means
that we have two isomorphic integrable systems with different
Lax representations and the corresponding canonical transformation
is a quasi-point transformation \cite{ts96a}.

To illustrate this construction we take as an example several
systems at $n=2$. Starting with an hyperelliptic curve $\C$ of
genus $g=n=2$ we define variables $(p_1,q_1)$ and $(p_2,q_2)$
on the Lagrangian submanifold $\C^{(2)}$ (\ref{sepv}). The
Jacobi inversion problem is the problem of finding these
variables from the equations (\ref{stinv}) with the St\"{a}ckel
matrix $\bs$ given by (\ref{stmd}). This problem is solved by
using the Kleinian $\wp$-functions, which are second
logarithmic derivatives of the Kleinian $\s$-function
\[
\wp_{ij}=-\dfrac{\p \ln\s(\b_1,\b_2)}{\p\b_i\p\b_j}\,,\qquad
\wp_{22}=q_1+q_2\,,\qquad \wp_{12}=-q_1q_2\,,
\]
(for detail see \cite{bak97,es96}).
The function $e(\l)$ (\ref{el}) on ${\C}$ with zeroes
at the points $q_1,q_2$ is equal to
\bq
e(\l)=\l^2-\wp_{22}\l-\wp_{12}=(\l-q_1)(\l-q_2)
=\l^2+2\l \sx_1+(2\sx_2 +\sx_1^2)\,,
\eq
or
\bq
e(\l)=\dfrac{(\l-q_1)(\l-q_2)}{(\l-\d_1)(\l-\d_2)}
=1+\dfrac{{\sx'}_1^2}{\l-\d_1}+\dfrac{{\sx'}_2^2}{\l-\d_2}\,.
\eq
Here we used the freedom (\ref{metrq}) and cartesian coordinates
$\{\sx_j\}$ or $\{{\sx'}_j\}$ are derived from the
"inverse" Lam\'e function $\B(\l)$ (\ref{bfl}).
Applying the outer additive automorphism of $sl(2)$, we
can construct the Lax matrices $L'(\l)$ for an infinite set of
integrable mechanical systems with the following hamiltonians
\ben
H&=&\sp_1\sp_2+V_N(\sx_1,\sx_2)\,,\nn\\
{\rm}\qquad&~&\ll{hhh}\\
H&=&{\sp'}_1^2+{\sp'}_2^2+V'_N(\sx'_1,\sx'_2)\,.\nn
\en
Among them, we distinguish the Henon-Heiles systems at $N=3$ and
the systems with quartic potential at $N=4$.  For these systems
the genus of associated curve $\C$ is equal to the number of
degrees of freedom $g=n=2$.

Function $e(\l)$ (\ref{aleq}) is independent on the moduli of
$\C$ and, therefore, the above construction of the integrable
systems (\ref{hhh}) readily gets over on the reducible
curve $\C$.  To construct this reducible curve, let us take two
tori $T_{1,2}^2$
\bq
w_{\pm}^2=\xi(1-\xi)(1-k_{\pm}^2\xi)\,,\ll{tori}
\eq
with a Jacobi moduli
\[k_{\pm}^2=-\dfrac{(\sqrt{\a}\mp\sqrt{\b})^2}{(1-\a)(1-\b)}\,.\]
Making the rational order two ($K=2$) change of variables
\bq
w_{\pm}=-\sqrt{(1-\a)(1-\b)}\dfrac{\l\mp
\sqrt{\a\b}} {(\l-\a)^2(\l-\b)^2}y\,,\qquad
\xi=\dfrac{(1-\a)(1-\b)}{(\l-\a)(\l-\b)}\l\,,\ll{cov}
\eq
one gets hyperelliptic curve
\bq
\C:\qquad y^2=\l(\l-1)(\l-\alpha)(\l-\beta)(\l-\alpha\beta)\,,
\ll{ccur}
\eq
which gives a two-sheeted covering of two tori $T^2_{1,2}$ (\ref{tori}).
It is a well-known example of the reduction of hyperelliptic
integrals to elliptic ones by using the rational change of
variables proposed by Legendre and generalized by Jacobi
\cite{bak97}.

The complex torus $T^2$ is isomorphic to the curve of genus
$g=1$ given by equation $w^2=f(\xi)$. In the
above, we have presented the covering for the two odd curves
(\ref{tori}) at deg$(f)=2g+1=3$. All computations concerning
the even curves at deg$(f)=2g+2=4$ give similar covering
\cite{bak97}, so we do not present these formulae.  The odd and
even curves at $g=1$ are associated to the integrable cases of
the Henon-Hailes system and system with quartic potential,
respectively.

Next we can introduce two pairs of variables
$(\te{p}_1,\te{q}_1)$ and $(\te{p}_2,\te{q}_2)$ being on the
tori $T^2_{1,2}$. Functions $e_{1,2}(\l)$ on $T^2_{1,2}$ are
equal to
\bq
e_1(\l)=\l-\te{q}_1\,,\qquad e_2(\l)=\l-\te{q}_2\,.\ll{2e}\eq
Variables $\{\te{p}_j,\te{q}_j\}$ are separated cartesian
coordinates for the integrable systems on $T^2_1\times T^2_2$
with the hamiltonians
\bq
H_{3,4}=\te{p}_1^2+\te{p}_2^2+V_{3,4}(\te{q_1})+V_{3,4}(\te{q}_2)
\,,\ll{hh}
\eq
which is a sum of two one-dimensional hamiltonians on $T^2_{1,2}$.
The corresponding $ 4\times 4$ Lax representation has a block
form (\ref{bllax}), whose blocks are determined by the functions
$e_{1,2}(\l)$ (\ref{2e}).

The covering (\ref{cov}) induces canonical transformation
of variables $\{\te{p}_j,\te{q}_j\}$ to $\{p_j,q_j\}$
\cite{es96}. These pairs of variables lie on the different
curves $T^2_{1,2}$ and $\C$, respectively.  The common moduli $\a$ and $\b$
of these curves are integrals of motion. On the orbit $\cal O$
($\a=const$, $\b=const$) this canonical transformation
(\ref{cov}) becomes a point transformation. It is so-called
quasi-point transformation \cite{ts96a}.

By using change of variables induced by covering
(\ref{cov}) one can construct
the $2\times 2$ Lax matrix for the evolution (\ref{hh})
splitting on two tori. In variables $\{\te{q}_j\}$ matrix
$L(\l)$ is determined by the function
\bq
e(\l)=\dfrac{(\l-\a)(\l-\b)}{(1-\a)(1-\b)}\te{e}(\l)\,,\qquad
\te{e}(\l)=(\l-\te{q}_1)(\l-\te{q}_2)\,.\ll{et}
\eq
In fact, we add two additional zeroes $\a$ and $\b$ into the
function $e(\l)$ (\ref{aleq}) on the reducible curve $\C$
(\ref{ccur}) and, therefore, change parameterization of the Lax
matrices in flat coordinates $\{\sp_j,\sx_j\}$.

In general, to introduce new flat coordinates,
we can take any tori $T^{2g_j}_{1,2}$ of arbitrary genus
$g_{1,2}>1$ and consider two-dimensional evolution (\ref{hh})
splitting on these curves with an arbitrary one-dimensional
potentials $V_{2g_j+1}(q_j)$. The standard change of
variables
\bq
\te{q_j}=\dfrac{\te{\sx}_1\pm\te{\sx}_2}2\,,\qquad
\Rightarrow\qquad
\te{e}(\l)=\l^2-\te{\sx}_1+\dfrac{\te{\sx}_1^2-
\te{\sx}_2^2}{4}\,,\ll{rep1}
\eq
preserves the natural form of the hamiltonians (\ref{hh})
for arbitrary potentials $V_{2g_j+1}(q_j)$. The equations of
motion remain the Newton equations in these variables
$\{\te{\sp}_j,\te{\sx_j}\}$

In the considered above example (\ref{tori}) both independent
hyperelliptic integrals are reduced to elliptic ones by using
a common substitution $\xi\to \l$ (\ref{cov}). It relates to
existence of the second order automorphism of a hyperelliptic curve
(\ref{ccur}) \cite{bak97}:
\bq
\tau:\qquad (\l,y)\to
\left(\dfrac{\a\b}{\l},\dfrac{y}{\l^3}\sqrt{\a^3\b^3}\right)\,.\ll{tau}
\eq
It allows us to introduce another parameterization of the
function $\te{e}(\l)$ in cartesian coordinates, which preserves
the natural form of the hamiltonian. Namely, in
addition to (\ref{rep1}), we can use the following canonical
transformation of variables $\{\te{p}_j,\te{q}_j\}$ to the
cartesian coordinates $\{\hat{\sp}_j,\hat{\sx}_j\}$
\bq
\te{e}(\l)=\l^2-\dfrac{Q_++Q_-}{\hat{\sx}_1}
+\dfrac{(Q_+-Q_-)^2}{4\hat{\sx}_1}\,.
\ll{rep2}
\eq
Here functions $Q_\pm(\hat{\sp}_j,\hat{\sx}_j)$ are the
classical counterparts of the supercharges in two-dimensional
SUSY \cite{ts96a} with the following properties
\[\{H,Q_\pm\}=\pm f(\hat{\sp}_j,
\hat{\sx}_j)Q_\pm\,,
\qquad
\{H,Q_+Q_-\}=0\,.
\]
At $g=2$ ($N=3$ or $N=4$ in (\ref{hh})) these functions
$Q_\pm$ and $f$ on variables $\{\te{p}_j,\te{q}_j\}$ or
$\{\hat{\sp}_j,\hat{\sx}_j\}$ are listed in \cite{ts96a}.
Moduli $\a$ and $\b$ in (\ref{tau}) are integrals of motion,
therefore, automorphism $\tau$ induces
a second quasi-point transformation associated to torus
$T_1^2\times T_2^2$ (\ref{tori}).

Two quasi-point transformations (\ref{cov}) and
(\ref{rep2}) for the physical variables $\{\sx'_j\}$, $\{\te{\sx}_j\}$
and $\{\hat{\sx}_j\}$ bind together all the integrable cases of the
Henon-Hailes system at $N=3$ and three integrable cases of the
system with quartic potential at $N=4$. Of course, these
systems have a common set of action-angle variables.  Moreover,
the same variables are associated to the Kowalewski top
\cite{du81}, which is a supersymmetric quantum model as well.

Thus, several supersymmetric models are related to evolution
splitting on the tori, when the number of degrees of freedom
is equal to the genus $g=n$ of the associated covering
curve $\C=T_1\times T_2$. It would be interesting to get a
geometrical interpretation of these supersymmetric objects
arising from finite-dimensional SUSY quantum mechanics.


\section{Conclusion}
\setcounter{equation}{0}
It is known, that curves $y+y^{-1}=F(\l)$ together with the
$1$-forms
\[dS^{(4)}=\l\dfrac{dy}{y}\,,\qquad
dS^{(5)}=\log\l\dfrac{dy}{y}\,,\]
are implied by integrable models of the Toda chain family
(standard and relativistic models).
The corresponding Lax representations are defined on the
Poisson-Lie groups with quadratic $r$-matrix algebra. The
corresponding mapping from the action-angle variables to
separated variables has been proposed in \cite{fm76}.

On the other hand we can consider the umbilic solutions of the
KdV equation \cite{almar92,alchm94}. These systems are defined on
a generalized Jacobi variety of the symmetric product of $n$
logarithmic Riemannian surfaces in place of the Liouville tori.
Nevertheless, it is possible to introduce variables that
linearize the corresponding hamiltonian flows.
These systems may be interpreted as counterparts of the
discrete-time St\"{a}ckel systems.

Both these sets of models are associated to the change of
parameterization of hyperelliptic curve from "plane"
parameterization to "annulus" ones ($\l\to\log\l$). The crucially
interesting lift to the interpretation of $\l$ as a coordinate on elliptic
curve.

For all these integrable models it would be interesting to
estimate the possibility of application of the usual
St\"{a}ckel approach. On this way we should consider
mapping between action-angle variables and
separated variables, and should study the differential of this
map. In the presented paper there are the Jacobi inversion
problem and differential of the Abel-Jacobi map.

This research has been partially supported
by RFBR grant 96-0100537.

\end{document}